\documentclass[11pt,twoside]{article}  

\usepackage{asp2006}  
\usepackage{epsf}  
\usepackage{psfig}  
\usepackage{lscape}  
\usepackage{graphicx}  
  
\begin{document}  
\title{Dissecting Massive YSOs with Mid-Infrared Interferometry}   
\author{H.~Linz\altaffilmark{1}, Th.~Henning\altaffilmark{1}, B.~Stecklum\altaffilmark{2},   
A.~Men'shchikov\altaffilmark{3}, R.~van Boekel\altaffilmark{1}, R.~Follert\altaffilmark{2},   
M.~Feldt\altaffilmark{1}}   
\altaffiltext{1}{Max--Planck--Institut f\"ur Astronomie Heidelberg, K\"onigstuhl 17, 69117 Heidelberg, Germany}  
\altaffiltext{2}{Th\"uringer Landessternwarte Tautenburg, Sternwarte 5, 07778 Tautenburg, Germany}  
\altaffiltext{3}{CEA Saclay/DSM/DAPNIA, Service d'Astrophysique, Orme des  
              Merisiers, Bat. 709, F-91191 Gif-sur-Yvette Cedex, France}   
  
\begin{abstract} 
  The very inner structure of massive YSOs is difficult to trace. With  
  conventional observational methods we often identify structures  
  still several hundreds of AU in size.  But we also need information  
  about the innermost regions where the actual mass transfer onto the  
  forming high-mass star occurs. An innovative way to probe these  
  scales is to utilise mid-infrared interferometry. Here, we present  
  first results of our MIDI GTO programme at the VLTI. We observed 10  
  well-known massive YSOs down to scales of 20 mas. We clearly resolve  
  these objects which results in low visibilities and sizes in the  
  order of 30\,--\,50~mas. Thus, with MIDI we can for the first time  
  quantify the extent of the thermal emission from the warm  
  circumstellar dust and thus calibrate existing concepts regarding  
  the compactness of such emission in the pre-UCH{\sc ii} region  
  phase. Special emphasis will be given to the BN-type object M8E-IR  
  where our modelling is most advanced and where there is indirect  
  evidence for a strongly bloated central star.  
  
\end{abstract}  
  
  
\section{Introduction}   
High-mass stars predominantly form in clustered environments much  
farther away from the Sun, on average, than typical well-investigated  
low-mass star-forming regions. Thus, high spatial resolution is a  
prerequisite for making progress in the observational study of  
high-mass star formation. Furthermore, all phases prior to the main  
sequence are usually deeply embedded.  This often forces observers of  
deeply embedded massive young stellar objects (MYSOs) to move to the  
mid-infrared (MIR) where the resolution of conventional imaging is  
limited to $>$ 0\farcs25 even with 8-m class telescopes. Hence, one  
traces linear scales still several hundred AU in size even for the  
nearest MYSOs, and conclusions on the geometry of the innermost  
circumstellar material remain ambiguous. MIR emission moderately  
resolved with single-dish telescopes could even arise from the inner  
outflow cones \citep[e.g.,][]{2006ApJ...642L..57D,2005A&A...429..903L}. \\  
The most versatile method to overcome the diffraction limit of single  
telescopes is to employ interferometric techniques. We are presently  
conducting a larger survey toward MYSOs based on MIR interferometry. In  
total, we have observed 10 sources so far.  All these sources, mostly  
comprising BN-type objects \citep[cf.][]{1990FCPh...14..321H}, are clearly  
resolved with the interferometer baselines we applied ($> 40$ m). This  
in itself is a major step forward compared to the previous more or  
less unresolved thermal infrared imaging with 4-m to 8-m class  
telescopes for these sources.  For this contribution, we concentrate  
on the object M8E-IR.  This is a prominent BN-type MYSO at a distance  
of roughly 1.5 kpc\footnote{Note, however, \citet{2007MNRAS.374.1253A}  
  for suggesting $d$ = 1.25 kpc for the M8 star clusters.} according  
to \citet{1984ApJ...278..170S}.  Although M8E-IR was a well  
investigated object in the 1980's, the spatial resolution for most of  
the IR observations of M8E-IR was limited. An exception is the work by  
\citet{1985ApJ...298..328S} who speculated on the existence of a small  
circumstellar disk around M8E-IR based on thermal infrared  
lunar occultation data. \\  
  
\section{MIDI interferometry for M8E--IR}\label{Sec:MIDI-Obs}  
  
Interferometric data in the mid-infrared wavelength range 8--13 $\mu$m  
have been obtained with the two-element interferometer MIDI  
\citep{2003SPIE.4838..893L} at the Very Large Telescope Interferometer  
(VLTI).  Within the framework of Guaranteed Time Observations (GTO)  
and Director's Discretionary Time (DDT) for MIDI, we observed M8E-IR at  
seven baseline length / baseline orientation combinations between June  
2004 and June 2005.  In Table \ref{Tab:MIDI-Log} we summarise the  
observations. We list the UT dates and times for the fringe track  
data, the projected baseline lengths and the position angles of the  
projected baselines on the sky (counted from north over east on the  
sky), as well as the used telescope configurations and the observing  
proposal numbers. We refer to \citet{2004A&A...423..537L} for a more  
detailed description of the standard observing procedure for MIDI  
observations. For all our observations, the so-called HighSens mode  
was used: during self-fringe tracking, all the incoming thermal  
infrared signal is used for beam combination and fringe tracking,  
while the spectro-photometry is subsequently obtained in separate  
observations. We used the MIDI prism as the dispersing element to get  
spectrally dispersed visibilities ($R \approx 30)$.  HD 169916 was  
used as the main interferometric and photometric standard star and was  
observed always immediately after M8E-IR. In addition, all calibrator  
measurements of a night were collected to create an average  
interferometric transfer function and to assign error margins to the  
measured visibilities. For the August 01, 2004 observations, the  
conditions were almost photometric, and the airmass during the  
observations of both, M8R-IR and HD 169916 was minimal (1.01).  
Therefore, we used the dispersed photometry from this measurement to  
provide the N-band spectrum later used in the SED fitting (see  
Sect.~\ref{Sec:M8E-IR-Modelling}).

\section{Results for M8E--IR}  
  
\begin{figure}[ht]  
\hspace*{-0.5cm}\includegraphics[width=7cm]{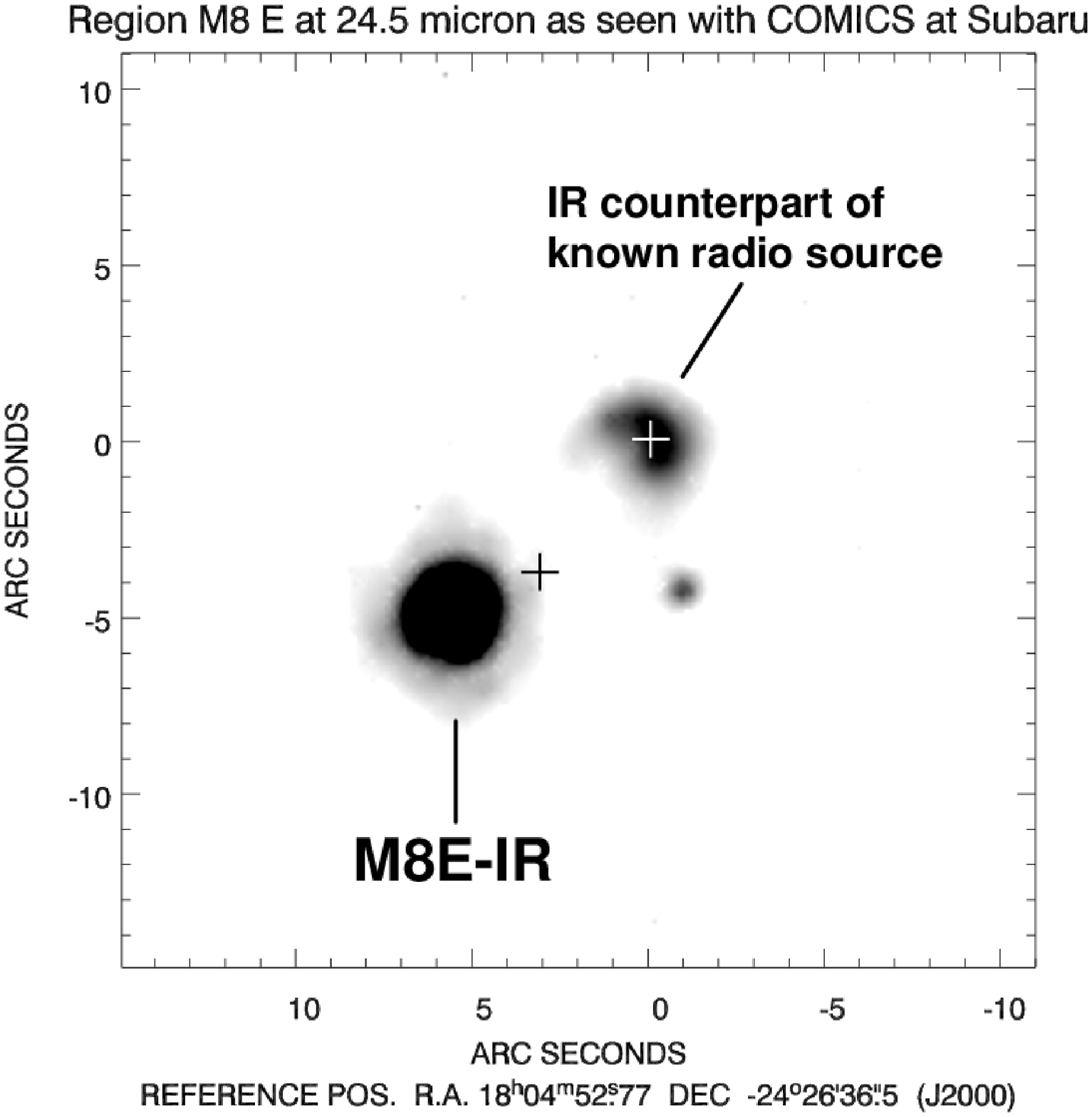}\includegraphics[width=7cm]{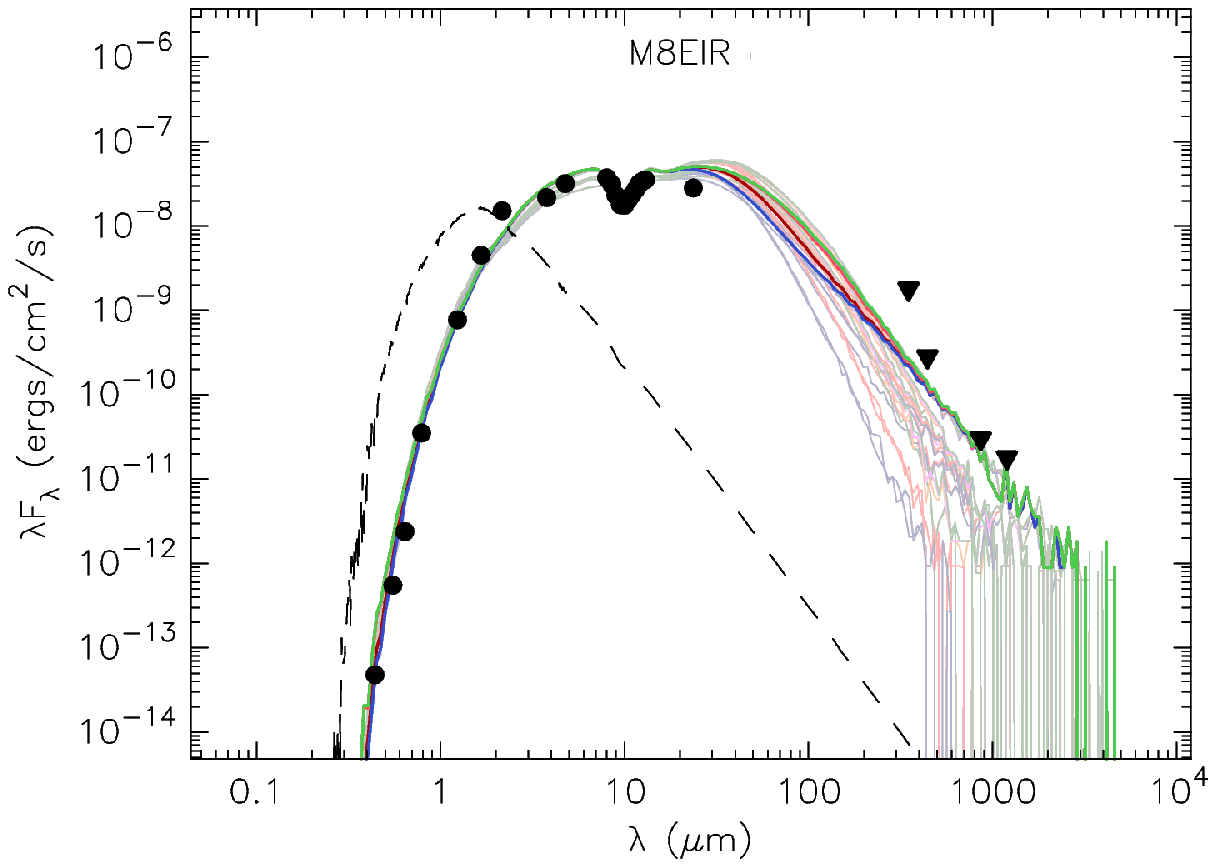}  
\caption{{\bf Left:} Q-band image of the M8E region at 24.5 $\mu$m with  
COMICS/Subaru. The two previously known objects are annotated.  
The black plus indicates the position of a 44 GHz Methanol maser  
according to \citet{1999ARep...43..149V}. The white plus  
marks the position of the cm continuum source   
\citep{1998A&A...336..339M}. {\bf Right:} SED of M8E-IR. Black dots show   
observed fluxes while black triangles show upper limits. The continuous lines give  
the best-fitting radiative transfer model from the grid of models by  
\citet{2007aApJS..169..328R} (different curves for different synthetic aperture sizes).  
The dash-dotted curve shows the SED of the bloated central star of that model.  
	   }  
      \label{Fig:Subaru}  
\end{figure}

To have a fresh look on M8E--IR in its environment, we obtained  
Subaru/CO\-MICS \citep{2003SPIE.4841..169O} thermal infrared imaging  
of this source from the SMOKA data archive  
\citep{2002ASPC..281..298B}.  We show the corresponding 24.5 $\mu$m  
Subaru/COMICS image in Fig.~\ref{Fig:Subaru} (Left).  M8E-IR is still  
the dominating source at this wavelength.  At a nominal resolution of  
0\farcs75, the emission remains compact. A second faint point source  
not yet reported in the literature is visible roughly 6$''$ west of  
it. Furthermore, we clearly detect MIR emission arising from the  
neighbouring radio source \citep{1984ApJ...278..170S,  
  1998A&A...336..339M}.  \citet{1985ApJ...298..328S} had already  
reported a detection of this source in N- and Q-band (fluxes only,  
measured with a 6$''$ diaphragm). For the first time, the Subaru MIR  
imaging spatially resolves this emission.  It is cometary-shaped, with  
the apex directed away from M8E-IR. This morphology could be an  
intrinsic property of this UCH{\sc ii} region, or it could be shaped  
by the molecular outflow probably (arising from M8E-IR,  
\citet{1988ApJ...327L..17M, 1992ApJ...386..604M}) influencing this radio  
source.  
At 10.5 $\mu$m we easily detect M8E-IR, but the radio source counterpart is not  
detected (rms noise $\approx$ 4 mJy). Considering the COMICS filter   
characteristics, combined with the fact that the   
radio counterpart was detected by \citet{1985ApJ...298..328S} with 0.97 Jy in a   
broad N-band filter, this implies a very strong and broad 9.7 $\mu$m Silicate   
absorption feature. The third source in the 24.5 $\mu$m image is detected at   
a $4 \sigma$ level at 10.5 $\mu$m.  
  
  
   \begin{table}  
      \hspace*{-0.25cm}       
      \caption[]{Log of MIDI observations of M8E-IR.}\label{Tab:MIDI-Log}  
    \begin{tabular}{lcccc}   
            \hline  
            \noalign{\smallskip}  
    UT \, date \, and \, time  &  B  &  PA   & Telescope & ESO \\  
                               & [m] & [deg] &  pair     & Program\\   
            \noalign{\smallskip}  
            \hline  
            \noalign{\smallskip}  
       2004-06-05 \, 08:06:57 & 96.8  &    +42.7  &  U1--U3  & 073.C-0175(A)\\  
       2004-06-05 \, 09:57:23 & 82.2  &    +44.5  &  U1--U3  & 073.C-0175(A)\\  
       2004-08-01 \, 01:50:41 & 46.6  &    +38.4  &  U2--U3  & 273.C-5044(A)\\  
       2005-03-02 \, 08:58:04 & 46.8  &  $-$85.9  &  U3--U4  & 074.C-0389(B)\\  
       2005-06-24 \, 07:36:50 & 51.6  &  $-$42.2  &  U3--U4  & 075.C-0755(B)\\  
       2005-06-24 \, 09:28:58 & 43.4  &  $-$10.6  &  U3--U4  & 075.C-0755(B)\\  
       2005-06-26 \, 00:26:11 & 55.7  &  $-$06.6  &  U1--U2  & 075.C-0755(A)\\  
            \noalign{\smallskip}  
            \hline  
         \end{tabular}  
   \end{table}  
%
\subsection{MIR interferometry}  
  
\begin{figure}[ht]  
\hspace*{-0.5cm}\includegraphics[width=7cm]{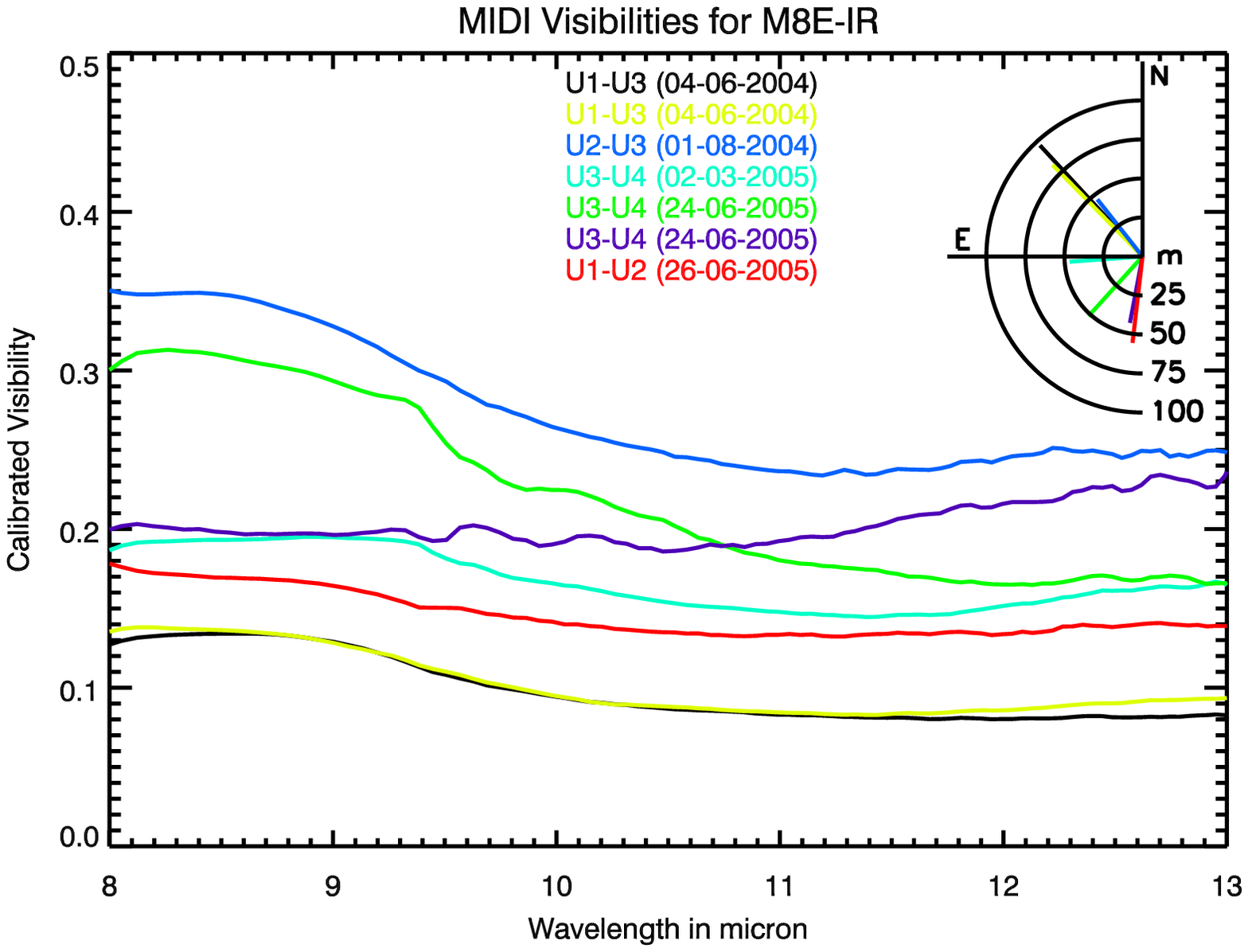}\includegraphics[width=7cm]{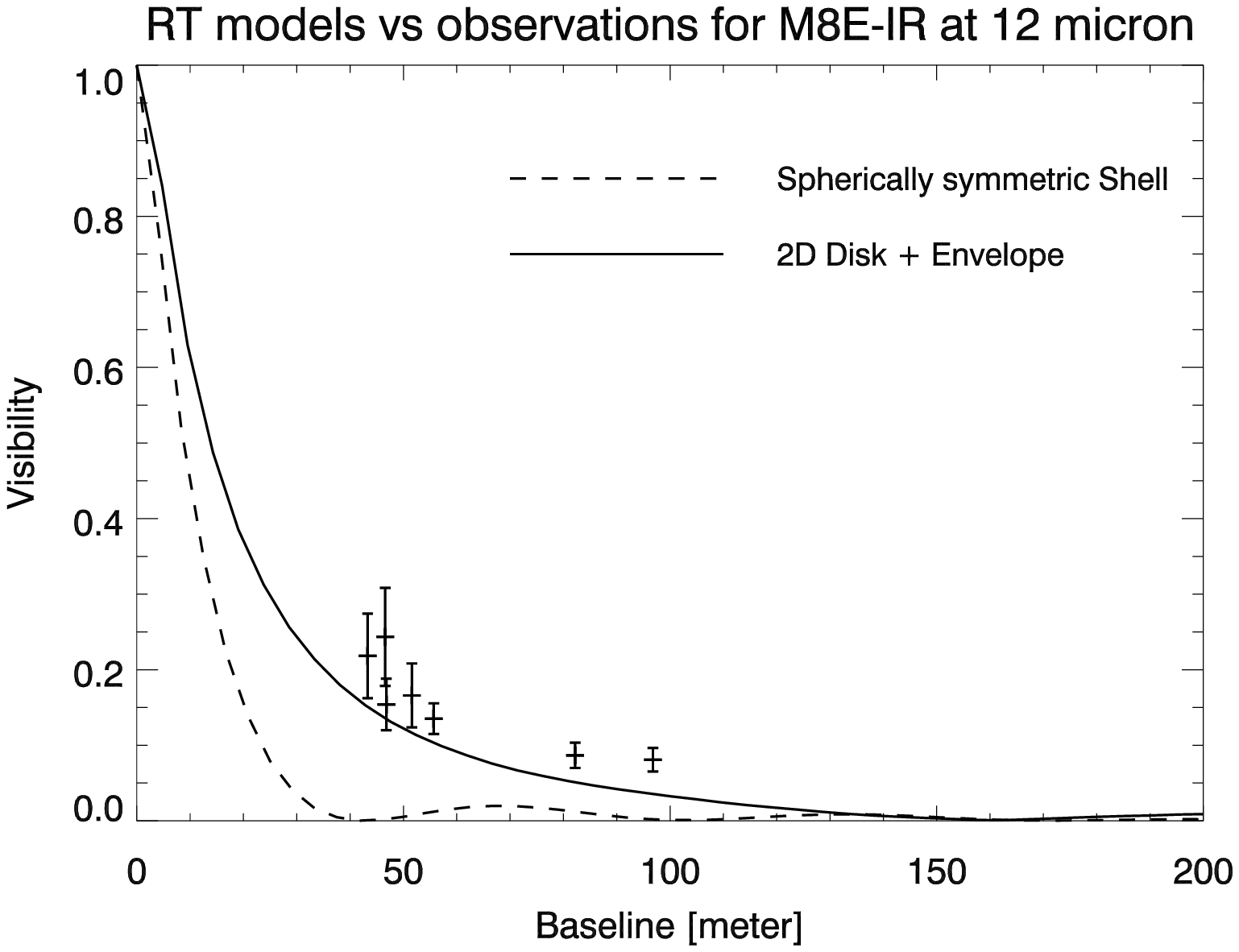}  
   \caption{{\bf Left:} Collection of dispersed MIDI visibilities. The inset  
schematically shows the baseline lengths and orientations   
(cf.~Table \ref{Tab:MIDI-Log}). {\bf Right:} Comparison of the spherically symmetric model   
incl.~a hot central star and the full 2D model with   
envelope + disk + bloated cool star. Given are cuts  
through the 12 $\mu$m model images with the resulting visibilities over  
the spatial frequencies (in units of the interferometric baseline length).  
MIDI data for the 7 baselines are also plotted as plus signes including   
the formal 3$\sigma$ error bars for comparison.}\label{Fig:RT-Comp}  
\end{figure}

We have reduced the interferometric data with the MIA+EWS package, version 1.5,  
developed at the MPIA Heidelberg and the University of Leiden. The resulting   
visibility curves are collected in Fig.~\ref{Fig:RT-Comp} (Left). Starting from these  
curves we can already make some statements. The object is clearly resolved in all   
our configurations with visibilities between 0.09--0.35. If we assume a Gaussian   
intensity distribution of the source  
the visibilities indicate an intensity FWHM of $\approx$ 20--25 mas  
(8.5 $\mu$m) and 32--38 mas (12.0 $\mu$m) which is in rough agreement  
with the extension of the small component of  
\cite{1985ApJ...298..328S}. We note, that these visibilities, although  
not reaching the relatively high levels of most Herbig Ae/Be stars  
\citep[e.g.,][Preibisch et al., these  
proceedings]{2004A&A...423..537L} seem to be qualitatively different  
from the very low visibilities (0.01\,--0.05) found for several of the  
other objects in our sample as well as recently reported for two other massive  
YSOs \citep[][Vehoff et al., these proceedings]{2007ApJ...671L.169D}.  
Still, to learn more about the potentially more complicated intensity  
structure, a simple Gaussian is not a sufficient ansatz (and physically  
not validated), especially if the object is strongly resolved by the  
interferometer.  Further modelling is necessary for interpretation.  
  
\subsection{Modelling}\label{Sec:M8E-IR-Modelling}  
  
We apply self-consistent continuum radiative transfer modelling to M8E-IR in   
order to produce synthetic MIR intensity maps and to compare their spatial   
frequency spectrum with the observed visibilities. Here, we are mainly concerned  
with the question which spatial distribution of the circumstellar material can   
account for both, the SED and the visibilities of M8E-IR. \\  
For SED fitting, we use the M8E-IR photometric data collected in   
\citet{2002aApJS..143..469M} plus new 1.2 mm data from \citet{2006A&A...447..221B}.   
We want to stress that no (sub-)millimeter interferometry   
on M8E-IR is reported in the literature which could spatially disentangle   
the flux contributions from M8E-IR and the radio source 8$''$ away. Hence, we   
consider the M8E-IR fluxes for $\lambda > 24.5 \, \mu$m just as upper limits in   
our modelling. Furthermore, we include new optical photometry in the B, V, and   
I filters reported by   
\citet[][their object Cl* NGC 6530 WFI 13458]{2005A&A...430..941P}   
as well as our new 24.5 $\mu$m photometry. In addition, the 8--13 $\mu$m total  
flux spectrum taken in the course of the MIDI measurements   
(Sect.~\ref{Sec:MIDI-Obs}) was used to further constrain the multitude of viable   
models. \\  
We started with well-tested models where the parameters just show a radial   
dependence \citep{1999ApJ...519..257M}. With a purely spherically symmetric geometry   
and a hot central star it was already possible to find models that  
reasonably fit the optical to MIR SED of M8E-IR. Furthermore, the  
modelling is capable of disentangling the principle contributions from  
M8E-IR and the neighbouring radio source to the total energy budget by  
assuming and modelling two distinct components. M8E-IR dominates up to  
wavelengths of 30\,--\,40 $\mu$m. To the total luminosity of around  
$1.7 - 2.0 \ \times \ 10^4$ L$_\odot$  
\citep[cf.~also][]{2002aApJS..143..469M}, M8E-IR might contribute less  
than 50\%.  Still, compared to the measured visibilities, these models  
resulted in far too low visibilities ($<$ 0.05) for M8E-IR in the  
baseline range 30--60 m over the whole 8--13 $\mu$m range. The error  
margins of the MIDI visibilities ($\approx\,10$\%) do not account for  
such large differences  
(Fig.~\ref{Fig:RT-Comp}). \\  
Hence, we extended our parameter search to full 2D models. We used the  
SED online fitting tool of \citet{2007aApJS..169..328R} to include  
models with envelope + circumstellar disk. We refer to this  
publication for details on the setup of these models. The best fits  
point to models comprising a very compact circumstellar disk ($<$ 50  
AU) and a larger surrounding envelope with small bipolar cavities. Due  
to the linkage of the Robitaille grid to evolutionary tracks, certain  
size parameters cannot be independently chosen by the fitter (see Robitaille,  
these proceedings). This especially affects the possible disk size which  
is not well constrained in our case. The data would still allow for a  
somewhat larger disk. Note however, that for disk radii clearly larger than  
100~AU, the MIR visibilities eventually drop. We mention explicitly, that  
among the fitting models there are also configurations without a disk  
(axisymmetric flattened envelope + outflow cavities only), which  
nevertheless give similarly high visibilities. This suggests that in  
the case of M8E--IR, the choice of the central object might actually  
govern the resulting visibility levels (see below). In  
Fig.~\ref{Fig:Subaru} (Right) we show the best SED fit to M8E-IR by  
the Robitaille models.  In Fig.~\ref{Fig:RT-Comp} (Left) the {\it  
  u,v}-spectrum of the 12 $\mu$m synthetic image based on this model  
is included as a solid line. It is compared to the corresponding image  
from the above-mentioned spherically symmetric modelling (dashed  
line).  Although the {\it u,v}-spectrum of the 2D model still shows  
somewhat lower visibilities than the measured ones, obviously it is  
qualitatively different from the spherically symmetric model. Due to  
the low inclination of the best 2D model (18$^\circ$), the central  
intensity distribution shows only small deviations from radial  
symmetry. Hence, differences in the model visibilities with varying  
position angle remain small ($<10$\%), and the cut shown in  
Fig.~\ref{Fig:RT-Comp} can be adopted as representative.  We  
explicitly mention the low inclination angle of $< 30^\circ$ inherent  
in all our 2D models that fit the SED data. This is in contrast to the  
disk hypothesis of \citet{1985ApJ...298..328S} that featured a very  
large inclination. However, Simon et al. saw a more or less symmetric  
intensity distribution in their 10 $\mu$m lunar occultation data;  
hints for deviations came mainly from lunar occultations recorded at  
3.8 $\mu$m.  At this shorter wavelength, configurations with disk +  
envelope incl.~outflow cones are naturally more structured, and  
scattered light in the inner regions of these cones can still  
contribute, in particular if grains larger than the  
typical 0.1 $\mu$m dust particles are involved. \\  
Interestingly, the best-fitting models for M8E-IR in the Robitaille  
model grid features central stars of 10--15 M$_\odot$ which are  
strongly bloated (120\,--\,150 R$_\odot$) and, therefore, have  
relatively low effective temperatures.  Such solutions can occur since  
the Robitaille grid also comprises the full range of canonical  
pre-main sequence evolutionary tracks from the Geneva  
group\footnote{{\tt http://obswww.unige.ch/$\sim$mowlavi/evol/stev\_database.html}}  
as possible parametrisation of the central objects. M8E-IR probably  
cannot straightforwardly be identified with correspondingly very  
early evolutionary stages, and we again refer to the contribution by  
Robitaille (these proceedings) for an extensive discussion on the  
intricate dependencies in his model grid. However, the tendency for a  
bloated central star in the case of M8E-IR may be valid. As mentioned  
already in \citet{1977A&A....54..539K}, accretion with high rates onto  
main sequence stars can temporarily puff up such stars 
\citep[see also][]{2007ARA&A..45..481Z}. Further encouragement to consider this  
comes from contributions presented during this conference (Hosokawa and  
Omukai as well as Yorke and Bodenheimer, these proceedings). These  
authors compute the pre-main-sequence evolution of stars in dependence  
of the accretion rate and find that for accretion rates (onto the  
forming star) reaching 10$^{-3} \, $M$_\odot$/yr, the protostellar  
radius can temporarily increase to $> 100$\,R$_\odot$, in accordance  
with our indirect findings from the model fitting.  Interestingly,  
\citet{1988ApJ...327L..17M} revealed high-velocity molecular outflows  
from M8E-IR based on M-band CO absorption spectra and speculated on  
recent ($<120$ yr) FU Ori-type outbursts for this object. If these  
multiple outflow components really trace recent strong accretion  
events, the central star could have indeed been affected.  
Furthermore, M8E-IR is not detected by cm observations with medium  
sensitivity \citep{1984ApJ...278..170S, 1998A&A...336..339M}. This  
could be explained by large accretion rates still quenching a forming  
hypercompact H{\sc ii} region \citep[e.g.,][]{1995aRMxAC...1..137W}.  
However, also a bloated central star with $T_{\rm eff} \ll 10000$ K  
would give a natural explanation for these findings.

\section{A new look onto M\,17 IRS1}  
  
\begin{figure}[ht]  
\hspace*{-0.0cm}\includegraphics[width=6.5cm]{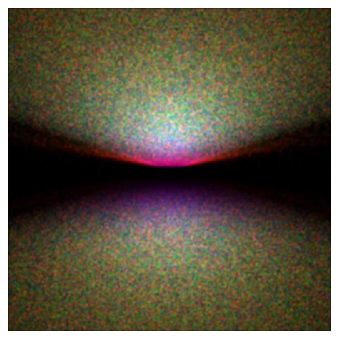}\hspace*{0.5cm}\includegraphics[width=6.5cm]{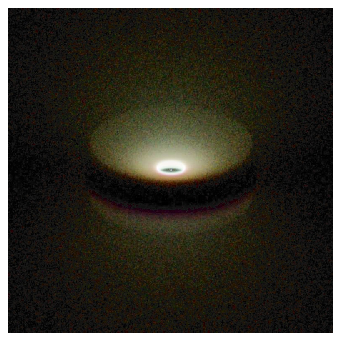}  
\caption{Synthetic 8\,--\,13 $\mu$m images (in logarithmic stretch) for the two best-fitting models   
delivered by the Robitaille fitter for M\,17 IRS1. Both images display a linear size of 2000 $\times$ 2000 AU$^2$.   
{\bf Left:} Nearly edge-on model with inclination angle 87$^\circ$, a disk radius of 2720 AU and a 17.5 M$_\odot$  
central star with $T_{\rm eff}$ = 33400 K.   
{\bf Right:} Model with inclination angle 63$^\circ$, a disk radius of 525 AU and a 7.9 M$_\odot$  
central star with $T_{\rm eff}$ = 15800 K. This model fits the measured visibilities much better than the  
other model.}\label{Fig:KWO}  
\end{figure}  
In order to give another example for the abilities of MIR  
interferometry to assess the viability of different geometric models  
for a massive YSO, we briefly mention our preliminary results for the  
object M\,17 IRS1 \citep[e.g.,][]{Chini}.  Here,  
we have obtained three visibility measurements with different position  
angles for baselines from 43~m to 56~m. We find clearly different  
visibility levels for the three measurements, staying at a 0.05-level  
for one measurement but reaching about 0.3 for the two other ones.  
Also here, we consulted the Robitaille SED fitter in order to find  
models reproducing the SED. Since the SED of M\,17 IRS1 is less  
constrained in the literature than in the case of M8E-IR, the fitter  
allows for a larger variety of models, all however including an  
intermediate-sized circumstellar disk. In Fig.~\ref{Fig:KWO} we show  
two of the best-fitting models. It gets obvious that in the 87$^\circ$  
edge-on model to the left we see relatively diffuse emission only,  
since the central source is hidden by the outer disk.  Consequently,  
this model results in far too low N-band visibilities and does  
{\it not} reach the measured visibility levels of 0.2\,--\,0.3 for any  
baseline orientation. The less inclined 63$^\circ$ model to the right  
at least provides a direct view onto the inner hot disk rim and  
therefore results in clearly higher visibilities than the edge-on  
model.

\section{Conclusion}    
We have observed massive young stellar objects with the  
MIR interferometer MIDI at the VLTI. We find substructures with MIR sizes   
around 30\,--\,50~mas. Using the measured visibilities as discriminator, we can exclude  
purely spherically symmetric matter distributions with a hot central star for   
the object M8E-IR. The most probably configuration consists of a compact   
circumstellar disk surrounded by a larger envelope, although the total disk size   
is not well constrained by the models. Until (sub-)millimeter observations with   
sufficient spatial resolution better constrain the emission at long wavelength,   
disks with radii from 20--100 AU seem probable. Furthermore, our data are 
consistent with the view that M8E-IR harbours a 10--15 M$_\odot$ central star 
that has been bloated by recent strong accretion events. Both aspects should be 
the topic of future investigations, for instance by utilising IR spectroscopy. \\  
For M\,17 IRS1, we can exclude the nearly edge-on disk model which had been suggested   
by the best-fitting 2D SED model. Instead, a moderately inclined circumstellar disk might be  
closer to the truth, and the results support an earlier suggestion that M\,17 IRS1 is a kind  
of still embedded Herbig Be star.\\  
Our results show that IR interferometry is a viable tool to reveal decisive 
structure information on embedded MYSOs and to resolve ambiguities arising from 
fitting the spectral energy distribution. With the inclusion of the auxiliary telescopes,
the VLTI is currently becoming even more flexible to tackle such tasks. Finally, 
2nd-generation VLTI instruments like {\sc Matisse} \citep{2006SPIE.6268E..31L} for the 
thermal IR will reveal even more complex details of MYSOs by adding imaging capabilities 
to MIR interferometry.

\acknowledgements 
Discussions with Th.~Robitaille, B.~Whitney, and\linebreak[4] M.~Hoare 
are appreciated. 
  

\end{document}